\documentclass[11pt,a4paper]{article}
\usepackage{graphicx}
\usepackage{amsmath}
\usepackage{amssymb}

\begin{document}

\title{CONTINUOUS-VARIABLE TELEPORTATION: \\ A NEW LOOK}

\author{ PAULINA MARIAN and TUDOR A. MARIAN\\
Centre for Advanced  Quantum Physics,\\
University of Bucharest, P.O.Box MG-11,\\
R-077125 Bucharest-M\u{a}gurele, Romania\\
E-mail: tudor.marian@g.unibuc.ro
}

\maketitle
\begin{center}
\end{center}

\begin{abstract}
In contrast to discrete-variable teleportation, a quantum state 
is imperfectly transferred from a sender to a remote receiver in a continuous-variable setting. We recall the ingenious scheme 
proposed by Braunstein and Kimble for teleporting a one-mode state 
of the quantum radiation field. By analyzing this protocol, we have previously proven the factorization of the characteristic function 
of the output state. This  indicates that teleportation 
is a noisy process that alters, to some extent, the input state.  
Teleportation with a two-mode Gaussian EPR state can be described 
in terms of the superposition of a distorting field with the input 
one. Here we analyze the one-mode Gaussian distorting-field state. 
Some of its most important properties are determined 
by the statistics of a positive EPR operator in the two-mode 
Gaussian resource state. We finally examine the fidelity 
of teleportation of a coherent state when using an arbitrary 
resource state.
\end{abstract}
 
{\centering\section{INTRODUCTION \label{int}}}

Quantum teleportation within continuous-variable (CV) settings 
is based on the same ideas as in the discrete case: 
these were put forward in the seminal work of Bennett 
{\em et al.} \cite{Ben}, who discovered the teleportation 
of qubits. The proposal of Braunstein and Kimble \cite{BK} 
was the first CV-teleportation scheme implemented experimentally. 
We find it useful to give here a succinct account 
of their protocol for teleporting a single-mode state 
of the quantum radiation field. 
 
Two distant operators, Alice, at a sending station, and Bob, 
at a receiving terminal, share an entangled two-mode state 
$\rho_{AB}$. Mode $A$ is operated by Alice and mode $B$ 
is controlled by Bob. When the cross-correlations between modes 
are strong enough, Alice and Bob can exploit the non-local 
character of the bipartite state $\rho_{AB}$ as a quantum 
resource for teleporting an unknown one-mode state $\rho_{in}$. 
The inseparable state $\rho_{AB}$ that connects the two parties 
is usually called an Einstein-Podolsky-Rosen (EPR) state. 
Without going into details, we briefly recall the successive 
steps of the  Braunstein-Kimble (BK) teleportation protocol. 

Alice performs a von Neumann measurement of a pair of commuting continuous variables. She combines two optical operations 
on her modes: mode mixing and homodyne detection. More specifically,
Alice mixes the input mode whose state $\rho_{in}$ is to be 
teleported with her $A$-mode of the EPR state $\rho_{AB}$ 
by employing a balanced lossless beam-splitter. As a result, 
she is ready to detect quadratures of the output modes.
By applying convenient projectors, Alice chooses to measure simultaneously the pair of commuting quadratures
\begin{equation}
\hat q_{A}:= \frac{1}{\sqrt2}(\hat q_{in}- \hat q_1),\;\;
 \hat p_{A}:= \frac{1}{\sqrt2}(\hat p_{in} +\hat p_1). \label{1.1}
\end{equation}
She conveys to Bob, through a {\em classical channel}, 
the result $\{q,p\}$ of her homodyne measurement as 
a complex amplitude, $\mu:=q+i p$. Any individual 
CV measurement performed by Alice is accompanied by a collapse 
of the initial tripartite state $\rho_{in} \otimes \rho_{AB}$. 
This results into a modified reduced $B$-mode state. Bob employs 
the value $\mu$ transmitted by Alice via the classical channel 
to perform a suitable displacement of the new reduced one-mode 
state at his side, $\rho_B(\mu)\rightarrow \rho^{\prime}_B(\mu)$.

The outcome $\mu$ is a continuous random variable. Therefore, 
Alice has to repeat her measurement under identical conditions 
in order to obtain a significant ensemble of results. She sends 
to Bob all these results, one by one, by successive classical communications. Every time, Bob operates the suitable displacement 
on the mode $B$ at his hand. He can thereby infer the distribution function ${\cal P}(\mu):={\cal P}(q,p)$ of the random 
variable $\mu.$ Bob is eventually able to build an imperfect 
replica of the initial state $\rho_{in}$ by averaging 
on the above-mentioned ensemble with the corresponding
distribution function: 
\begin{equation}
\rho_{out}=\int {\rm d}^2 \mu {\cal P}(\mu)\rho^{\prime}_B(\mu).
\label{1.4}
\end{equation}
Here we have denoted ${\rm d}^2 \mu:={\rm d}q \;{\rm d}p.$

Had we summarized the key steps of the BK protocol, 
this is explained in the framework of quantum mechanics, 
in Section 2, in terms of measurements and operations. 
Our main tool is the Weyl expansion of the density operators 
of the states involved. For subsequent use, we introduce
a non-local positive operator that we call EPR operator. 
Section 3 deals with two-mode Gaussian EPR states. We first
prove the existence of a one-mode {\em distorting-field state}
that is entirely determined by the EPR state. Its properties 
are then carefully analyzed and we show that a {\em Gaussian 
teleportation channel} does exist. The accuracy of the CV teleportation, measured either by the amount of added noise 
or by the fidelity of teleporting a coherent state, 
is investigated in Section 4.

\vspace{4mm}

{\centering\section{QUANTUM-MECHANICAL DESCRIPTION \label{QM} }}

\vspace{2mm}
\setcounter{equation}{0}
We present the BK protocol in the Schr\"odinger picture. The initial three-mode state  is the product $\rho_{in}\otimes \rho_{AB}$
of the one-mode state to be teleported,
\begin{eqnarray}
\rho_{in}=\frac{1}{\pi}\int {\rm d}^2 \lambda \;\;\chi_{in} (\lambda)
D(-\lambda), \label{cf1}
\end{eqnarray}
and of the two-mode EPR state,
\begin{eqnarray}
\rho_{AB}=\frac{1}{\pi^2}\int {\rm d}^2 \lambda_1 {\rm d}^2 
\lambda_2\;\;
\chi_{AB}(\lambda_1,\lambda_2)D_1(-\lambda_1)D_2(-\lambda_2).
\label{cf2}
\end{eqnarray}
Equations~(\ref{cf1}) and~(\ref{cf2}) exhibit the Weyl expansions 
of the corresponding density operators. We have denoted here 
by $D(\alpha):=\exp{(\alpha {\hat a}^{\dag}-\alpha^*{\hat a})}$  
a Weyl displacement operator on a single-mode Hilbert space: 
$\hat a$ is the mode annihilation operator. The states ~(\ref{cf1}) 
and ~(\ref{cf2}) are described in terms of their 
characteristic functions (CFs) $\chi_{in}(\lambda)$ and 
$\chi_{AB}(\lambda_1,\lambda_2)$, 
which are particular cases of the multimode definition 
\begin{equation}
\chi(\lambda_1,\lambda_2,\cdots ,\lambda_n):={\rm Tr}
[\rho D(\lambda_1)D(\lambda_2)\cdots D(\lambda_n)]. \label{cfs}
\end{equation}
The compatible observables ~(\ref{1.1}) have continuous spectra. 
Their common eigenfunction for an outcome $\{q,p\}$ 
of the homodyne measurement, 
\begin{eqnarray}
|\Phi(q,p)\rangle=\frac{1}{\sqrt{\pi}}
\int\limits_{-\infty}^{\infty}{\rm d}\eta{\rm e}^{i\sqrt{2} p\eta}
|\sqrt{2}q+\eta\rangle_{in}\otimes|\eta\rangle_A  \label{eigen},
\end{eqnarray}
satisfies the orthonormality condition
$$\langle \Phi(q^{\prime},p^{\prime})|\Phi(q,p)\rangle=
\delta (q^{\prime}-q)\delta (p^{\prime}-p).$$ 
The distribution function of the continuous random variable 
$\mu$ is
\begin{equation}
{\cal P}(q,p)={\rm Tr}_{in,AB}\left[M(\mu)\right], \label{df}
\end{equation}
where $M(\mu)$ is an operator on the Hilbert space 
${\cal H}_{in} \otimes {\cal H}_A \otimes {\cal H}_B$:  
\begin{equation}
M(\mu):=\left[|\Phi(q,p)\rangle
\langle\Phi(q,p)|\otimes I_B\right](\rho_{in}\otimes\rho_{AB})
\label{M(mu)}.
\end{equation}
As a result of the projective measurement performed by Alice, 
the initial product state $\rho_{in}\otimes \rho_{AB}$
collapses, so that the after-collapse $B$-mode reduction 
$\rho_B(\mu)$ can be written by tracing out the three-mode 
operator ~(\ref{M(mu)}) on the Hilbert space 
${\cal H}_{in} \otimes {\cal H}_A:$ 
\begin{eqnarray}
\rho_B(\mu)=\frac{1}{{\cal P}(\mu)}
{\rm Tr}_{in,A}\left[M(\mu)\right]. \label{rhoB}
\end{eqnarray}
The information provided by Alice allows Bob to perform a suitable displacement, 
$\rho_B^{\prime}(\mu)=D_2(\mu)\rho_B(\mu)
D^{\dag}_2(\mu)$. Then the ensemble averaging~(\ref{1.4}) 
yields the state emerging by CV teleportation from the unknown
input state $\rho_{in}$:
\begin{equation} 
\rho_{out}=\int {\rm d}^2 \mu \;{\cal P}(\mu)\;D_2(\mu)\rho_B(\mu)
D^{\dag}_2(\mu). \label{rhoout}
\end{equation}
Note that, owing to the CV-teleportation protocol itself,
the output single-mode state $\rho_{out}$, eq.~(\ref{rhoout}), 
is always a mixed one.

Our main previous result is a very simple formula connecting 
the one-mode states $\rho_{in}$ and $\rho_{out}$. It is
expressed in terms of their normally-ordered CFs and a remnant 
of the CF $\chi_{AB}(\lambda_1,\lambda_2)$ of the EPR state 
\cite{PT03,PT06,PT08}:
\begin{equation}
\chi_{out}^{(N)}(\lambda)= \chi_{in}^{(N)}(\lambda)
\chi_{AB}(\lambda^*,\lambda). \label{chinchi} 
\end{equation} 
The factorization formula ~(\ref{chinchi}) shows that 
CV teleportation is a noisy process, which always alters 
the input state $\rho_{in}.$ It is equivalent to the
identity
\begin{equation}
\chi_{out}(\lambda)= \chi_{in}(\lambda)
\chi_{AB}(\lambda^*,\lambda). \label{chichi} 
\end{equation} 
The function 
\begin{equation}
\chi_{D}^{(N)}(\lambda):=\chi_{AB}(\lambda^*,\lambda)
\label{chind}
\end{equation} 
leads us to introduce the related one,
\begin{equation}
\chi_{D}(\lambda):=\exp{\left (-\frac{1}{2}|\lambda|^2 \right )}
\chi_{D}^{(N)}(\lambda), \label{chid}
\end{equation}
which enters the one-mode Weyl expansion
\begin{equation}
\rho_D:=\frac{1}{\pi}\int {\rm d}^2 \lambda \;\;\chi_{D} (\lambda)
D_2(-\lambda). \label{rod}
\end{equation}
Here  $\rho_D$ is a self-adjoint Hilbert-Schmidt operator 
of unit trace, on the single-mode Hilbert space ${\cal H}_B$ 
at Bob's side.

Unless the two-mode EPR state $\rho_{AB}$ is Gaussian, one could 
only conjecture the positivity of the operator $\rho_D$. However, 
in the Gaussian case, we will prove this property in the next 
section. Accordingly, for the special class of the two-mode 
Gaussian EPR states, the function $\chi_{D}(\lambda)$, eq.~(\ref{chid}), is the CF of a $B$-mode state $\rho_D$ that we have termed 
{\em distorting-field state} \cite{PT08}. 
The multiplication rule ~(\ref{chinchi}) displays therefore
the fact that $\rho_{out}$ is the $B$-mode state 
of a superposition of two single-mode fields: the input one 
and a remote distorting field in the state $\rho_D$, 
due to the imperfect character of CV teleportation.

We conclude this section by introducing some operators 
on the Hilbert space ${\cal H}_A \otimes {\cal H}_B$ 
that prove to be useful when analyzing the non-locality 
features of the two-mode resource state $\rho_{AB}$.
The usual EPR observables are two commuting linear combinations 
of the single-mode canonical operators: the {\em relative
coordinate} $\hat Q:= \hat q_1-\hat q_2$ and the {\em total
momentum} $\hat P:=\hat p_1+\hat p_2$.  
In terms of them we define the EPR operator
\begin{equation}
\hat \Delta:=\frac{1}{2}\left({\hat Q}^2+{\hat P}^2\right). 
\label{d} 
\end{equation}
For later convenience, let us introduce a non-local normal 
operator,
\begin{eqnarray}
\hat A:=\hat a_1-(\hat a_2)^{\dag}=\frac{1}{\sqrt{2}} 
(\hat Q+i \hat P), \label{A}
\end{eqnarray}
as a suitable amplitude of the positive EPR operator: 
\begin{equation}
\hat \Delta={\hat A}^{\dag} \hat A.
\label{a} 
\end{equation}

\vspace{4mm}

{\centering\section{GAUSSIAN DISTORTING-FIELD STATE}}

\vspace{2mm}
\setcounter{equation}{0}
In what follows we adopt a shorthand notation concerning 
the operator $\rho_D$, eq.~(\ref{rod}): we denote by ${\cal H}$
its one-mode  Hilbert space and by $\hat a$ the corresponding
annihilation operator. Moreover, in order to simplify 
the subsequent discussion, we assume for the moment 
that a distorting-field state $\rho_D$ does exist 
whatever EPR state $\rho_{AB}$. We eventually prove that 
this assumption is true for any two-mode Gaussian EPR state.

To start on our analysis, substitution into eq.~(\ref{chind})
of the Taylor expansions 
\begin{eqnarray}
\chi^{(N)}_D(\lambda)=\sum_{l, m=0}^{\infty}
\frac{1}{l!m!}
\lambda^{l}(-\lambda^{*})^{m}\langle({\hat a}^{\dag})^{l}
{\hat a}^{m}\rangle_D
\label{chin}
\end{eqnarray}
and
\begin{eqnarray}
\chi_{AB}(\lambda^*,\lambda)=\sum_{l, m=0}^{\infty}
\frac{1}{l!m!}
\lambda^{l}(-\lambda^{*})^{m}\langle({\hat A}^{\dag})^{l}
{\hat A}^{m}\rangle_{AB}
\label{ab}
\end{eqnarray}
yields the correlation functions in the distorting-field state: 
\begin{eqnarray}
\langle({\hat a}^{\dag})^{l}{\hat a}^{m}\rangle_D\
=\langle({\hat A}^{\dag})^{l}{\hat A}^{m}\rangle_{AB}.
\label{cfs1}
\end{eqnarray}
We will omit subsequently the pair of indices $AB$ when 
writing expectation values in the EPR state $\rho_{AB}$:
$$\langle\ldots\rangle:=\langle\ldots\rangle_{AB}.$$
In particular, the $l$th-order correlation function 
is non-negative for any $l$:
\begin{equation}
\langle({\hat a}^{\dag})^{l}{\hat a}^{l}\rangle_D\
=\langle{\hat \Delta}^{l}\rangle \geq 0.
\label{dcf}
\end{equation}
The identity ~(\ref{cfs1}) can be employed to evaluate
the $2\times 2$  covariance matrix (CM) 
of the distorting-field state $\rho_D$ \cite{PTH01}:
\begin{eqnarray}
{\cal V}_D=\left(\begin{array}{cc}\sigma_D(q, q)&\sigma_D(q, p)\\
\sigma_D(p, q)&\sigma_D(p, p)\end{array}\right)
\label{vm}
\end{eqnarray} 
Explicitly, the CM ${\cal V}_D$  has the following entries 
\cite{PT08}:
\begin{eqnarray}
\sigma_D(q, q)=\frac{1}{2}+\langle{\hat Q}^2\rangle,\;\;\;
\sigma_D(q, p)=\langle\hat Q \hat P\rangle,\;\;\;
\sigma_D(p, p)=\frac{1}{2}+\langle{\hat P}^2\rangle. \label{ent}
\end{eqnarray}
Owing to the Schwarz inequality for a quasi-inner product,
\begin{equation}
{\langle\hat Q \hat P\rangle}^2 \leq 
\langle \hat Q^2\rangle  \langle \hat P^2\rangle,
\label{Schwarz} 
\end{equation}
the Robertson-Schr\"odinger uncertainty relation holds:
\begin{equation}
{\cal V}_D+\frac{i}{2}J\geq 0, \;\;\;(J:=i\sigma_2),  
\label{RS} 
\end{equation}
with $\sigma_2$ denoting the complex Pauli matrix.
Condition ~(\ref{RS}) is necessary for all single-mode states,
but it is sufficient only for the Gaussian ones.
By virtue of definition ~(\ref{chind}), the function 
$\chi_{D}(\lambda)$, eq.~(\ref{chid}), is Gaussian 
if and only if the two-mode EPR state $\rho_{AB}$ is Gaussian too.
In this case, $\chi_{D}(\lambda)$ is indeed the CF of a one-mode Gaussian state $\rho_D$: our assertion is therefore proven. 

From now on, in this section, we deal only with Gaussian CV teleportation. The existence of a Gaussian distorting-field 
state $\rho_D$ allows us to read eq.~(\ref{chinchi}) 
as a multiplication rule of normally-ordered CFs: 
\begin{equation}
\chi_{out}^{(N)}(\lambda)= \chi_{in}^{(N)}(\lambda)
\chi_{D}^{(N)}(\lambda). \label{chinchin} 
\end{equation} 
Note that if the Gaussian resource state $\rho_{AB}$ 
is undisplaced, so is the distorting-field state $\rho_D$. 
Further, according to eqs.~(\ref{ent}), $\rho_D$ is a mixed 
state, since $\det{\cal V}_D >1/4$, unless the random 
EPR variables $\hat Q$ and $\hat P$ are constants, 
meeting thus the ideal EPR demand. 

Moreover, eqs.~(\ref{ent}) show that a matrix inequality 
holds, ${\cal V}_D \geq \frac{1}{2}I_2,$ pointing out 
that $\rho_D$ is a {\em classical} state. Therefore, 
the Gaussian distorting-field state $\rho_D$ 
has a regular Glauber-Sudarshan $P$ representation,
\begin{equation}   
P_D(\alpha)=\frac{1}{\pi}\int{\rm d}^{2}\lambda   
\exp{(\alpha \lambda^*-\alpha^* \lambda)} 
\chi_D^{(N)}(\lambda), \label{p}
\end{equation}
which is a Gaussian distribution function. 
The multiplication rule ~(\ref{chinchin}) is equivalent 
to the existence of a mapping
\begin{equation}   
\rho_{out}=\int{\rm d}^{2}\beta P_{D}(\beta)D(\beta)
\rho_{in}D^{\dag}(\beta) \label{channel}
\end{equation}
between one-mode states on the Hilbert space ${\cal H}$.  
We call such a mapping a {\em teleportation channel}. 
In sum, Gaussian CV teleportation is described 
by a Gaussian channel ~(\ref{channel}). 

It is instructive to evaluate the Glauber $R$ function 
of the state $\rho_D$ \cite{Glauber},
\begin{equation}   
R_D(\beta^*, \beta^{\prime}):=\exp{\left(\frac{1}{2}(|\beta|^2
+|\beta^{\prime}|^2)\right)}\langle\beta |\rho_D|
\beta^{\prime}\rangle, \label{r}
\end{equation}
as an integral \cite{PT93}:
\begin{equation}   
R_D(\beta^*, \beta^{\prime})=\exp(\beta^*\beta^{\prime})
\frac{1}{\pi}\int{\rm d}^{2}\lambda \chi_D^{(N)}(\lambda) 
\exp{(-|\lambda|^2-\beta^* \lambda+\beta^{\prime} 
\lambda^*)}. \label{rint}
\end{equation}
Making use of the eqs.~(\ref{chin}) and ~(\ref{cfs}), we find:
\begin{equation}   
R_D(\beta^*, \beta^{\prime})=\langle\exp{(-{\hat \Delta}
+\beta^{\prime}{\hat A}^{\dag}+\beta^*{\hat A})}\rangle. 
\label{rd}
\end{equation}
It follows that the Husimi function ($\equiv$ the Glauber 
$Q$ function) reads:
\begin{equation}   
Q_D(\beta):=\frac{1}{\pi}\langle\beta |\rho_D|\beta\rangle 
=\frac{1}{\pi}\exp(-|\beta|^2)\langle\exp{(-{\hat \Delta}
+\beta{\hat A}^{\dag}+\beta^*{\hat A}})\rangle. \label{qd}
\end{equation}
We take advantage of the Taylor expansion of the $R$ function 
\cite{Glauber},
\begin{equation}   
R_D(\beta^*, \beta^{\prime})=\sum_{l, m=0}^{\infty}
(\rho_D)_{lm}\frac{1}{\sqrt{l!m!}}(\beta^*)^{l}
(\beta^{\prime})^{m}, \label{rser}
\end{equation}
to write down the density matrix
\begin{equation}   
(\rho_D)_{lm}=\frac{1}{\sqrt{l!m!}}\langle\exp{(-{\hat \Delta)}}
{\hat A}^{l}({\hat A}^{\dag})^{m}\rangle. \label{dm}
\end{equation}
The corresponding photon-number distribution, 
\begin{equation}   
(\rho_D)_{ll}=\frac{1}{l!}\langle{\hat \Delta}^{l}
\exp(-{\hat \Delta)}\rangle, \label{pnd}
\end{equation}
has the generating function 
$G_D(s):=\sum_{l=0}^{\infty}s^l(\rho_D)_{ll},\;(|s|\leq 1)$:
\begin{equation}   
G_D(s)=\langle\exp((s-1){\hat \Delta})\rangle. \label{gen}
\end{equation}
This distribution is entirely determined by the statistics 
of the EPR operator ${\hat \Delta}$, eq.~(\ref{d}), 
in the two-mode Gaussian EPR state $\rho_{AB}$. 
It is worth mentioning that eqs.~(\ref{chinchin})--~(\ref{gen}) 
hold also for any classical {\em non-Gaussian} 
distorting-field state $\rho_D$: in fact, classicality 
is required only to ensure the validity of eqs.~(\ref{p})
and ~(\ref{channel}), playing no role for the other ones.
 
Let us specialize the above discussion to a zero-mean Gaussian 
resource state frequently used, namely, a two-mode squeezed 
vacuum state (SVS): 
$\rho_{AB}=|\Psi_{AB}\rangle\langle\Psi_{AB}|.$  
The Schmidt decomposition of such a pure state 
in the standard Fock basis, 
$$|\Psi_{AB}\rangle=\frac{1}{\cosh r}\sum_{n=0}^{\infty} 
(\tanh r)^n |n\rangle_A \otimes |n\rangle_B,$$
is parametrized with the squeezing factor $r>0$. The corresponding distorting-field state is thermal, 
\begin{equation}  
\chi_{D}^{(N)}(\lambda)=\exp{(-{\rm e}^{-2 r}|\lambda|^2)}, 
\label{chiSVS} 
\end{equation}
and has the $P$ representation 
\begin{equation}
{P}_D(\alpha)=\frac{{\rm e}^{2 r}}{\pi}
\exp{(-{\rm e}^{2 r}|\alpha|^2)}.
\label{PSVS} 
\end{equation}
With eqs.~(\ref{chin}) and ~(\ref{cfs1}), we get 
the correlation functions 
\begin{eqnarray}
\langle({\hat a}^{\dag})^{l}{\hat a}^{m}\rangle_D\
={\delta}_{lm}l!({\rm e}^{-2 r})^l
=\langle({\hat A}^{\dag})^{l}{\hat A}^{m}\rangle.
\label{cfsSVS}
\end{eqnarray}
Note the mean photon number, 
$\langle {\hat a}^{\dag} {\hat a} \rangle_D=\exp{(-2 r)}
=\langle{\hat \Delta}\rangle$, 
and the $l$th-order correlation function, 
$\langle ({\hat a}^{\dag})^l {\hat a}^l \rangle_D
=l!{\langle{\hat \Delta}\rangle}^l
=\langle{\hat \Delta}^l\rangle$. 

For the sake of completeness, we write down further 
the $R$ function, 
\begin{equation}   
R_D(\beta^*, \beta^{\prime})=\frac{1}{1+{\rm e}^{-2 r}}
\exp{\left(\frac{{\rm e}^{-2 r}}{1+{\rm e}^{-2 r}}
\beta^* \beta^{\prime}\right)}, \label{R}
\end{equation}
the Husimi function,
\begin{equation}   
Q_D(\beta)=\frac{1}{\pi}\frac{1}{1+{\rm e}^{-2 r}}
\exp{\left(-\frac{|\beta|^2}{1+{\rm e}^{-2 r}}\right)}, 
\label{QSVS}
\end{equation}
the density matrix,
\begin{equation}   
(\rho_D)_{lm}={\delta}_{lm}\frac{1}{1+{\rm e}^{-2 r}}
\left(\frac{{\rm e}^{-2 r}}{1+{\rm e}^{-2 r}}\right)^l, 
\label{dmSVS}
\end{equation}
and the generating function of the photon-number distribution,
\begin{equation}   
G_D(s)=\left[1+(1-s){\rm e}^{-2 r}\right]^{-1}. \label{genf}
\end{equation}
All the above formulae are specific for a single-mode thermal state. 
Therefore, when using a SVS as the resource state, 
teleportation is described by a thermalization channel.

\vspace{4mm}

{\centering\section{ACCURACY OF TELEPORTATION}}

\vspace{2mm}
\setcounter{equation}{0}

Originally, the quality of the teleportation protocol 
was quantified  by the input-output overlap for pure states 
\cite{BK}, or by use of the Uhlmann fidelity for mixed 
Gaussian states \cite{PT03,Ban2}. For a clear survey 
on the progress in CV teleportation we refer the reader 
to Ref. \cite{PM}. More recently \cite{PT06,PT08}, in analyzing 
CV teleportation, the present authors have introduced 
the distorting-field state $\rho_D$ and focused on its properties. 
We point out here the conspicuous role of the EPR operator 
${\hat \Delta}$, eq.~(\ref{d}). Its expectation value 
in the resource state $\rho_{AB}$, called EPR uncertainty 
\cite{Duan}, quantifies the non-locality of this state.  
As the EPR uncertainty $\langle {\hat \Delta} \rangle$ decreases, 
the non-local character of the two-mode state $\rho_{AB}$ becomes stronger. In particular, the inequality 
$\langle {\hat \Delta} \rangle < 1$ is a criterion 
of inseparability of the bipartite state $\rho_{AB}$. 

We start by assuming first the existence of a one-mode 
remote-field state $\rho_D$: we have shown that this
effectively happens {\em at least} for the class 
of the two-mode Gaussian EPR states. The  quality 
of teleportation can be evaluated in terms of the mean 
photon number $\langle {\hat a}^{\dag}{\hat a} \rangle_D$
in the one-mode state $\rho_D$. For any undisplaced Gaussian 
EPR state, this can be seen as the amount of noise added 
by teleportation: it distorts the features of the input 
field state $\rho_{in}$. The smaller this noise, the higher 
the quality of the CV teleportation. According to eq.~(\ref{cfs1}), 
the added noise is equal to the EPR uncertainty:
\begin{equation}
\langle {\hat a}^{\dag} {\hat a} \rangle_D
=\langle {\hat \Delta} \rangle. \label{noise} 
\end{equation}

Second, we make a conjecture that extends this result to 
an arbitrary undisplaced two-mode EPR state:
{\em The amount of noise distorting the properties of 
the input field state is equal to the EPR uncertainty
$\langle {\hat \Delta} \rangle$}. 

A remarkable theorem proven by Giedke {\em et al.} \cite{G}
states that among all equally entangled pure two-mode states, 
the SVS has the minimal EPR uncertainty 
$\langle {\hat \Delta} \rangle$, 
{\em i.e.}, the strongest non-local character. 
This theorem regarding the ranking of pure-state 
entanglement at a given EPR uncertainty enables us 
to notice an interesting property of CV teleportation:
{\em The SVS adds the minimal noise in teleportation with 
pure two-mode resource states having the same entanglement.}

We finally give a new expression of another quantity 
that is widely employed to measure the teleportation accuracy: 
the fidelity of teleporting a coherent state, hereafter 
denoted by ${\cal F}_{coh}$. Recall that a coherent state 
$\rho_{in}=|\alpha\rangle \langle \alpha|$ has a Gaussian CF: 
\begin{equation}
\chi_{in}(\lambda)=\exp{(\alpha^* \lambda-\alpha \lambda^*)}
\exp{\left (-\frac{1}{2}|\lambda|^2 \right )}.
\end{equation} 
The fidelity of teleporting a coherent state is the  probability 
of the transition $\rho_{in} \rightarrow \rho_{out}:$ 
\begin{equation}
{\cal F}_{coh}=\langle \alpha|\rho_{out}|\alpha \rangle. 
\end{equation}
When writing this quantity in terms of the CFs of the states 
involved, eq.~(\ref{chichi}) provides an expression 
that is independent of the input coherent state:
\begin{equation}
{\cal F}_{coh}=\frac{1}{\pi} \int{\rm d}^{2}\lambda
\exp{(-|\lambda|^2)}\chi_{AB}(\lambda^*,\lambda).
\end{equation}
Making use of the Taylor series~(\ref{ab}), we find: 
\begin{equation}
{\cal F}_{coh}=\sum_{l=0}^{\infty}\frac{(-1)^l}{l!}
\langle{\hat \Delta}^l\rangle=\langle 
\exp{(-\hat \Delta)}\rangle. \label{sum}
\end{equation}
Equation~(\ref{sum}) is valid for any two-mode EPR state 
$\rho_{AB}$. It displays to what extent the EPR operator 
$\hat \Delta$ is involved in the structure of the fidelity of teleporting a coherent state. If a distorting-field state 
$\rho_D$ exists, then an inspection of eqs.~(\ref{qd})
and~(\ref{sum}) gives the identity
\begin{equation}
{\cal F}_{coh}=\pi Q_D(0). \label{fcoh}
\end{equation}
For instance, let us consider again the case of a SVS 
chosen as a two-mode resource state. Then, by use of 
eqs.~(\ref{QSVS}) and~(\ref{fcoh}), we recover the formula
\begin{equation}
{\cal F}_{coh}=\frac{1}{1+\exp{(-2 r)}}, \label{svs}
\end{equation}
in agreement with previous results \cite{BK,Ban2}.

To sum up, in this paper we have examined further 
the CF description of the BK teleportation protocol. 
For the class of the two-mode Gaussian EPR states, 
we have been able to identify a remote distorting 
field mode superposed on the input one. This originates 
in the noisy character of the CV teleportation. 
We have pointed out the main properties of a one-mode 
Gaussian distorting-field state. They are connected 
with the non-local features of the two-mode EPR state.  
The accuracy of the CV teleportation is measured either 
by the amount of added noise or by the fidelity 
of teleporting a coherent state. Both quantities depend 
on the degree of non-locality of the bipartite resource 
state expressed in terms of the EPR uncertainty.

\vspace{4mm}

{\centering\subsubsection*{Acknowledgements}}

\vspace{4mm}

This work was supported by the Romanian 
Ministry of Education and Research through Grant IDEI-995/2007 for the University of Bucharest.

\vspace{4mm}

\begin{center}

\end{center}
\end{document}